\documentclass[12pt,preprint]{aastex63}
\usepackage{amsmath}
\newcommand\kmps{\mbox{km\,s$^{-1}$}}
\newcommand\Msun{\mbox{$M_\sun$}}
\newcommand\Mspy{\mbox{\Msun\,yr$^{-1}$}}

\newcommand\rev[1]{{\bf #1}}

\shorttitle{Stellar Wind Accretion in Symbiotic Stars}
\shortauthors{Lee, Kim \& Lee}

\begin{document}
\title{Formation of the Asymmetric Accretion Disk from Stellar Wind Accretion in an S-type Symbiotic Star}

\author{Young-Min Lee}
\affiliation{Department of Physics and Astronomy, Sejong University, Seoul, 05006, Korea}
\affiliation{SEP Engineering, Anyang, Gyeonggi, 14059, Korea}

\author{Hyosun Kim}
\affiliation{Korea Astronomy and Space Science Institute, 776, Daedeokdae-ro, Yuseong-gu, Daejeon 34055, Republic of Korea}

\author{Hee-Won Lee}
\affiliation{Department of Physics and Astronomy, Sejong University, Seoul, 05006, Korea}

\correspondingauthor{Hyosun Kim}
\email{hkim@kasi.re.kr}

\begin{abstract}
The accretion process in a typical S-type symbiotic star, targeting~AG Draconis, is investigated through 3D hydrodynamical simulations using the {\tt FLASH} code.
Regardless of the wind velocity of the giant star, an accretion disk surrounding the white dwarf is always formed. In the wind models faster than the orbital velocity of the white dwarf, the disk size and accretion rate are consistent with the predictions under the Bondi-Hoyle-Lyttleton (BHL) condition. In slower wind models, unlike the BHL predictions, the disk size does not grow and the accretion rate increases to a considerably higher level, up to $>20\%$ of the mass-loss rate of the giant star.
The accretion disk in our fiducial model is characterized by a flared disk with a radius of 0.16~au and a scale height of 0.03~au. The disk mass of $\sim5\times10^{-8}\ \Msun$ is asymmetrically distributed with the density peak toward the giant star, being about 50\% higher than the density minimum in the disk.
Two inflowing spiral features are clearly identified and their relevance to the azimuthal asymmetry of disk is pointed out.
The flow in the accretion disk is found to be sub-Keplerian with about 90\% of the Keplerian speed, which indicates the caveat of overestimating the \ion{O}{6} emission region from spectroscopy of Raman-scattered \ion{O}{6} features at 6825\,\AA\ and 7082\,\AA.
\end{abstract}


\section{Introduction}\label{sec:intro}

Binary systems involving an accreting white dwarf (WD) are mainly classified into cataclysmic variables and symbiotic stars. Cataclysmic variables are semi-detached binary systems of an accreting WD and a late-type main-sequence star. In these stellar systems, the formation of an accretion disk through Roche lobe overflow has been well established. The main feature that characterizes cataclysmic variables is dwarf nova outbursts, which are attributed to siginificant change in viscosity depending on the surface mass density in the disk \citep[e.g.,][]{warner95}.

In contrast, a symbiotic star is a wide binary system consisting of a late-type evolved star with a large mass loss rate ranging $10^{-8}$--$10^{-6}\ \Mspy$ (\citealp{dupree86, hofner18, seaquist93}) and a hot accreting star, usually a WD \citep{kenyon86}. Symbiotic stars are classified into S-type and D-type based on their spectral energy distributions. D-type symbiotic stars exhibit infrared (IR) excess indicative of the presence of warm dust component with $T\sim 10^3$~K whereas no such IR excess appears in S-type objects. The orbital periods of S-type symbiotic stars are typically a few years whereas those of D-type symbiotics are poorly known and suggested to be in the range of several decades \citep{belczynski00}. 

The large binary separation of most symbiotic stars implies that the Roche lobe of the giant component is underfilled so that accretion onto the WD component occurs through gravitational capture of the slow stellar wind from the giant component.
\citet{muerset99} suggested that, in at least most symbiotic systems, the radius of the giant star is only half the distance to the inner Lagrangian point, which is empirically found from the strong correlation between the spectral types of cool giants and their orbital periods.
However, in some cases with short orbital period systems, mass transfer through the inner Lagrangian point may not be completely excluded \citep[e.g.,][]{boffin14}.

The luminosity of the hot component in symbiotic stars during the quiescent phase is found to be in the range $10^2$--$10^3\ L_\sun$, which is contributed mostly by stable thermonuclear reaction on the surface of the WD with an accretion rate of a few times $10^{-8}$--$10^{-7}\ \Mspy$, depending on the mass of WD \citep[see e.g., Fig.~2 of][]{shen07}. Major outbursts with the luminosity of order $10^4\ L_\sun$ are often attributed to thermonuclear burning  as a result of a substantially increased accretion rate of $\sim10^{-6}\ \Mspy$ \citep{sokoloski06, mikolajewska12}. The variability exhibited by symbiotic stars is apparently influenced by many factors including pulsations of the giant star, instability associated with the accretion flow and nonsteady thermonuclear burning on the surface of the WD \citep{sokoloski06}. 
 
A highly interesting spectroscopic tool to probe the accretion processes in symbiotic stars is provided from broad emission features at 6830\,\AA\ and 7088\,\AA\ known to be present in about a half of symbiotic stars \citep{allen80, akras19}. These spectral features are formed through inelastic scattering of \ion{O}{6} $\lambda\lambda$1032 and 1038 with atomic hydrogen \citep{schmid89}. They exhibit double or triple peak profiles with enhanced red peak and the peak separation is about 20--40\ \kmps. \citet{schmid96} carried out a pioneering radiative transfer study using the Monte Carlo technique to investigate the basic properties of Raman-scattered \ion{O}{6} features including the Raman scattering efficiency and the polarization structures. 

\citet{lee07} proposed that multiple-peak profiles are explained by invoking a Keplerian accretion disk of a peak separation of $\sim50\ \kmps$ with asymmetric density distribution augmented by modulation of stellar wind terminal velocity $\sim20\ \kmps$ and the presence of receding bipolar components with respect to the binary orbital plane. The geometry of neutral medium is complex and the accretion flow around the WD tends to be asymmetric, based on the previous Raman \ion{O}{6} spectroscopic investigations \citep[see][]{lee97b, heo15, lee19}.

\citet{lee19}, however, pointed out that the line profile analysis of Raman-scattered \ion{O}{6} has its limitation as the adopted model for the \ion{O}{6} emission region is purely kinematical with no proper dynamical considerations. Therefore, hydrodynamical studies of the stellar wind accretion in symbiotic stars will allow a quantitative estimation of asymmetric distribution of \ion{O}{6} emitters, leading to a more reliable description of the mass transfer through Raman spectroscopy than the purely kinematic model. 

The stellar wind accretion processes in a relatively close binary systems, corresponding to typical S-type symbiotic stars, are investigated by \citet{chen17, saladino18, saladino19a}. Their hydrodynamic simulations for a variety of orbital and wind parameters, including the binary separation of 3--20~au, clarified that the comparable speeds between the wind flow of the giant star and the orbital motion of the accretor (the companion star; the secondary star; the WD in a symbiotic star system) provide favorable conditions for disk formation. These studies focused on showing the effects of the orbital and wind parameters on the overall morphology of the matter surrounding the binary, the mass-accretion efficiency onto the accretor, the angular momentum loss of the binary orbit, and the change in the mass-loss rate of the giant star. Relatively less focused issues in these previous investigations include the physical structure of accretion disks, for example, density and velocity distributions, which constitute the main scope of this paper.

Some details of the disk formation by capturing wind material in binaries, albeit only for relatively wide systems with the separations of 10--20~au, are explored by \citet{huarte13}. Based on Bondi-Hoyle-Lyttleton (BHL) accretion theory \citep{HL39, BH44}, \citet{huarte13} derived the upper limit for the radius of the wind-captured disk surrounding the accretor as a function of the masses of the component stars, the orbital separation, and the wind velocity. This formula plays an important role to yield minimum resolution criteria for numerical simulations to properly resolve the wind capture time scale which is associated with the simulation grid resolution. In this earlier work, the disks occupied more than $\sim50$ cells within the radii for different models. Besides such an analytic treatment for the disk formation, their major results through numerical simulations include the discovery of the vortex tube-like accretion stream, and the measurement of the temporal change of the accretion disk in its mass and eccentricity. They provided new insights into the BHL accretion and a comprehensive analysis on the accretion streams in the binary systems. The formation process and the physical structure of a disk in a relatively closer binary, however, would be different; for instance, as noted by \citet{huarte13}, stripping of disk material due to the powerful ram pressure of the stellar wind and much higher accretion rates are expected in closer binaries.

In this work, we carry out 3D hydrodynamic simulations of stellar wind accretion processes in relatively close binaries mimicking typical S-type symbiotic stars. To understand the formation and evolution of accretion disks in close binaries, we present a quantitative analysis of the flows toward, and within, the accretion disks. In Section~\ref{sec:setup}, we introduce numerical methods adopted in this work. In Section~\ref{sec:res}, our fiducial model presents the gas flows forming a large-scale spiral structure surrounding the binary stars, of which the head interacts with the flows building an accretion disk around the WD. Several models by adopting different wind parameters are also exhibited in order to describe the formation criteria and time evolution of the accretion disks in this section. The observational ramifications for the S-type symbiotic star AG Dra are briefly described in Section~\ref{sec:obs}. Finally, we discuss the effect of the accretion and structure of the disks in Section~\ref{sec:discussion}.

\section{Simulation setup}\label{sec:setup}
\subsection{Equations of hydrodynamics}

In order to simulate the wind accretion processes in S-type symbiotic stars, we adopt the adaptive mesh refinement code {\tt FLASH} version 4.5 \citep{fryxell00} with modifications in the way similar to the works carried out by \citet{kim12b, kim13, kim17, kim19}. In this work, the fluid is assumed to be inviscid and non-self gravitating. The Eulerian hydrodynamic equations are used in the center of mass frame with the Cartesian coordinates.

Based on the continuity equation, 
\begin{equation}
    \frac{\partial\rho}{\partial t} + \nabla\cdot(\rho {\bf v}) = 0, 
\end{equation}
the initial density distribution is set by
\begin{equation}
  \rho_0 ({\bf r}) = \frac{\dot M_{\rm G}}
      {4\pi |{\bf r}-{\bf r}_{\rm G}|^2\, V_{\rm ej}},
\end{equation}
assuming a steady state stellar wind originated from the intrinsically spherically symmetric mass loss of a non-pulsating and non-rotating giant star situated at ${\bf r} = {\bf r}_{\rm G}$. The initial density distribution of the background is relaxed within an orbit of binary in the simulation.
Here, the parameter $V_{\rm ej}$ indicates the stellar wind velocity at $|{\bf r}-{\bf r}_{\rm G}|=r_*$ that we define as the surface of the giant star. The radius $r_*$, at which the wind density and velocity conditions are reset every simulation step, is set to 0.25 au, similar to the photospheric radius of the evolved giant star of S-type symbiotic systems (\citealp{skopal05}; see also \citealp{dumm98}). Because the radius of giant star is far less than the distance to the inner Lagrangian point ($\sim 1$~au) in the binary system employed in our simulations, the mass transfer through the Roche-lobe overflow does not occur. 
The adopted mass-loss rate of the giant star is $\dot M_{\rm G}=10^{-7}\ \Mspy$. Because the mass loss of the giant star during the total simulation time is negligible in terms of orbital evolution of the binary system, simplification is made by fixing the stellar masses and their orbits throughout the simulation.

The masses of the two stars are fixed to $M_{\rm G}=1.5\ \Msun$ and $M_{\rm WD}=0.6\ \Msun$. The two stars are orbiting around the center of mass of the binary system, following the assumed circular trajectories with the fixed separation of 2~au $(=|{\bf r}_{\rm G}-{\bf r}_{\rm WD}|)$. The corresponding orbital period is $\sim2$~yr. The orbital velocity of the WD is $V_{\rm WD}=21.8\ \kmps$ with respect to the center of mass of binary, and the relative velocity between the stars is $V_{\rm rel}=30.5\ \kmps$.
In Table~\ref{tab:repmod}, we summarize the fixed parameters adopted for the models exhibited in this paper.

The acceleration force on the wind material is determined by the equation of motion, 
\begin{equation}
  \frac{\partial{\bf v}}{\partial t} + ({\bf v}\cdot\nabla){\bf v} =
  -\frac{1}{\rho}\nabla P -\nabla\Phi_{\rm G}^{\rm eff} -\nabla\Phi_{\rm WD}.
\end{equation}
An adiabatic equation of state with the specific heat ratio $\gamma=5/3$ for a monatomic ideal gas is used to update the pressure $P$. We also set the effective temperature of the giant star, $T_{\rm eff}=4,000$~K, which is proper for the giant star of S-type symbiotic stars as presented by \citet{skopal05}.

The specific gravitational forces $-\nabla \Phi_{\rm G}^{\rm eff}$ and $-\nabla \Phi_{\rm WD}$ due to the presence of the giant star and WD are defined by the Plummer models with the softening radii $r_{\rm soft, G}$ and $r_{\rm soft, WD}$, respectively \citep{bt08}:
\begin{equation}
\begin{split}
  -\nabla\Phi_{\rm G}^{\rm eff} = -\frac{GM_{\rm G}(1-f)}
  {|{\bf r}-{\bf r}_{\rm G}|^2 + r_{\rm soft, G}^2}, \\ \nonumber 
  -\nabla\Phi_{\rm WD}          = -\frac{GM_{\rm WD}}
  {|{\bf r}-{\bf r}_{\rm WD}|^2 + r_{\rm soft, WD}^2}.
\end{split}
\end{equation}
Here, $G$, $M_{\rm G, WD}$ and ${\bf r}_{\rm G, WD}$ are the gravitational constant, the individual masses of the giant star and the WD, and the position vectors of the stars in the center of mass coordinates, respectively. The softening radius $r_{\rm soft, G}$ for the giant star is set to 0.25 au, being comparable to the expected photospheric radius of the giant star (see above). The size of the WD (order of $10^{-5}$~au) cannot be resolved in our simulations because of the limitation of computation time, therefore we set the softening radius $r_{\rm soft, WD}$ to be the smallest length scale at a given resolution of a simulation model, within which we require more than the minimum 4 grid cells (see Table~\ref{tab:totmodel}). The acceleration factor $f$ of unity is assumed in all of our simulations \citep[see Section~2.1 of][for details of the wind model]{kim12a}.

The orbital velocity of the WD relative to the wind velocity {\it in situ} is an important factor regulating the detailed properties of accretion disk. The simulations are carried out by varying the intrinsic wind velocity, ${V_{\rm w}}$, defined as the velocity at the distance to the WD (i.e., 2~au) but in the corresponding single-star simulation, in which the WD is intentionally absent. Alternatively speaking, $V_{\rm w}$ is determined merely by the balance between the gas pressure and the gravitational potential of the giant star including the acceleration factor mimicking the radiation pressure onto the hypothetical dust particles. The velocity $V_{\rm w}$ for each model is listed in Table~\ref{tab:totmodel}. 

\subsection{Treatment for cooling}

In the energy equation, we include the radiative cooling as two separate terms following \citet{saladino18}:
\begin{equation}\label{eq:energy}
  \frac{\partial E}{\partial t} + \nabla\cdot[(E+P){\bf v}] =
  -\rho \left[ \frac{3k}{2\mu m_{\rm p}} \frac{(T-T_{\rm eq})\rho}{C} + \dot Q_{\rm sh} \right].
\end{equation}
Here, $E$, $T$, $P$, ${\bf v}$, and $\rho$ are the basic quantities that describe the fluid, representing the total energy, temperature, pressure, velocity vector, and mass density, respectively. The constants $k$ and $m_{\rm p}$ are the Boltzmann constant and proton mass, respectively, and $\mu=1.29$ is the mean molecular weight appropriate to the solar metalicity \citep[e.g.,][]{anders89}.

The first cooling term on the right hand side of Eq.~(\ref{eq:energy}) regulates the gas temperature being balanced with the equilibrium temperature \citep{chandrasekhar34}, 
\begin{equation}\label{eq:tau}
    T_{\rm eq} = T_{\rm eff} (W+3/4\tau^\prime)^{1/4},   
\end{equation}
within the equilibrium time scale, $C/\rho$, where the constant $C = 10^{-5}\rm\ g\,s\,cm^{-3}$ is derived from an approximate density-dependent expression of radiative cooling rate in the atmosphere of giant stars \citep{bowen88}. The effective temperature $T_{\rm eff}$ represents the temperature at the photosphere of the giant star. The parameter $W = (1/2)[1-(1-r_*^2/s^2)^{1/2}]$ indicates the geometrical dilution factor at a distance $s=|{\bf r}-{\bf r}_{\rm G}|$ from the giant star, while the optical depth $\tau^\prime = \int_s^\infty (\kappa_{\rm g}+\kappa_{\rm d})\, \rho\, r_*^2/s^2\ ds$  and the factor 3/4 in Eq.~(\ref{eq:tau}) are determined assuming the plane-parallel atmosphere. The opacities for gas and dust, $\kappa_{\rm g}$ and $\kappa_{\rm d}$, are $2\times10^{-4}$ and $5\rm\ cm^2\,g^{-1}$, respectively. 
Notice that when the gas temperature drops below $T_{\rm eq}$, this first cooling term behaves like a heating source in regard to the radiation of the giant star, preventing unrealistic reduction of the gas temperature.

The second term on the right hand side of Eq.~(\ref{eq:energy}) describes the radiative cooling taking into account neutral hydrogen and heavier chemical elements in a gaseous medium with the solar abundance. The cooling rate is defined as
\begin{equation}
    \dot Q_{\rm sh} = \frac{\Lambda_{\rm hd} n_{\rm H}^2}{\rho},
\end{equation}
with the cooling function $\Lambda_{\rm hd}$ provided by \citet{schure09} in their Table~2. The hydrogen number density $n_{\rm H} = X \rho /m_{\rm p}$ is calculated with the fixed mass fraction of hydrogen in the gas ($X=0.707$). 
When the gas temperature rapidly increases at the shock fronts that are established at the spiral arms or at the vicinity of the WD under the strong gravitational potential well, the second cooling term plays an important role in efficiently reducing the internal energy of the gas. It therefore provides a favorable condition for disk formation surrounding the WD.
It should be noted that the low temperature regime ($\log_{10} T < 3.8$) is not considered in this radiative cooling term, and thus the gas temperature in such a regime stays near the equilibrium temperature as determined by the first cooling term in Eq.~(\ref{eq:energy}).

\subsection{Accretion sink}

The WD is treated as an accretion sink with the sink radius $r_{\rm sink}$ in order to prevent unrealistic penetration of flows through the WD and to obtain legitimate values for the accretion rate. In our simulation, the radius $r_{\rm sink}$, at which the accretion sink occurs, is set to be the same as the gravitational softening radius $r_{\rm soft, WD}$. When the flow approaches to the WD within the radius $r_{\rm sink}$ and its energy condition satisfies the gravitationally bound state, the flow is regarded as being entered the sink and no further trace is made. Instead, the density within the sink radius is replaced by an arbitrary small value, the sink density $\rho_{\rm sink}$, that is predefined to be much smaller than the background density, as practically the {\it in-situ} density in the absence of the WD. The velocity within the sink radius is replaced by the orbital velocity of the WD.
The accretion rate, $\dot M_{\rm acc}=\sum_i \rho_i\, dU_i / dt$, is recorded at the end of each simulation timestep of the time duration $dt$, where the density and volume elements of the $i$-th grid cell within the accretion sink, satisfying the sink condition, are denoted by $\rho_i$ and $dU_i$.
Several tests have been performed to verify that our choice for $\rho_{\rm sink}$ does not aﬀect the overall morphology of accretion ﬂows and their physical properties. We have also performed tests to confirm the sufficiency of numerical resolution of assigning at least 4 cells within the accretion sink radius \citep[see also][]{truelove97}. In tests for sink radius reduced by half (accompanied by proper grid resolution), negligible changes are observed in the disk radius and height measurements as well as the density and temperature ranges within the accretion disk.

\subsection{System coordinates}

The computational domain is set to be a rectangular cube having the dimensions of $12.8\times12.8\times6.4$ au$^3$ with the center of mass of the binary stars $(x=y=z=0)$ at the center of the first and second coordinates but on the boundary of the third coordinate, assuming symmetry about the equatorial plane ($z=0$). 
The boundary condition for the plane coinciding with the equatorial plane is to reflect all quantities in the $z>0$ space to the $z<0$ space through the mirror at the $z=0$ plane. The boundary conditions for the other five boundary planes are set to {\tt diode}, defined in the {\tt FLASH} code, that allows outflow but prohibits inflow of fluid.
At each simulation time step, the meshes for high density regions are refined through the adaptive mesh refinement technique with the predefined maximum refinement level of 6 (or level 7 in one model for resolution test) and $64\times64\times32$ cells being included in each level of refinement.
To interpolate physical quantities of each grid cell, the Piecewise Parabolic Method is adopted in this work.

In analyses for the accretion disk of the WD, we introduce the coordinates $(x^\prime,\,y^\prime,\,z^\prime)$ defined in the rest frame of the WD. 
We also note that the notations $r$ and $R$ indicate the radii in the spherical and cylindrical coordinates, respectively.
The angle $\Theta$ is defined about the WD in the counterclockwise direction starting from the opposite side of the location of the giant star.

\section{Results}\label{sec:res}
\subsection{Simulation models and overall morphology}\label{sec:simmod}

We have carried out eight simulations Sim1--Sim8 with different wind conditions, and their input parameters and the resulting properties of accretion disks are tabulated in Table~\ref{tab:totmodel}.
Our experiments are designed to investigate the physical properties of the outflows and the accretion disks surrounding the WD, according to the increase of $V_{\rm w}$ from Sim1 to Sim6.
The intrinsic wind velocity of Sim1, measured from an unperturbed single-star simulation, starts from $V_{\rm ej}=6\ \kmps$ at the surface of the giant star (i.e., at radius $r_*$) and reaches $V_{\rm w}=13.7\ \kmps$ at the distance corresponding to the binary separation. This wind model mimics a transonic wind, as the sound speed at the surface of the giant star is $\sim7.4\ \kmps$. The winds of the giant star in the other seven models are intrinsically supersonic. 
A slow wind with $V_{\rm w}$ smaller than the orbital velocity $V_{\rm WD}=21.8\ \kmps$ (Sim1--Sim5) facilitates the capture of stellar wind material, making a favorable condition for the formation of an accretion disk surrounding the WD.

A fast wind with the velocity $V_{\rm w}$ exceeding the orbital velocity $V_{\rm WD}$ tends to pass over the WD instead of effectively building a disk around it. Under such an unfavorable condition for disk formation, the Sim6 model failed in creating a disk. With reduced sink radii, the Sim7 and Sim8 models are computed in order to assess the effect of the sink radius and the grid resolution on the formation of accretion disk. We assure our minimum grid criterion, 4 cells within radius $r_{\rm sink}$, for these models by increasing the maximum refinement level.

\begin{figure*} 
  \plotone{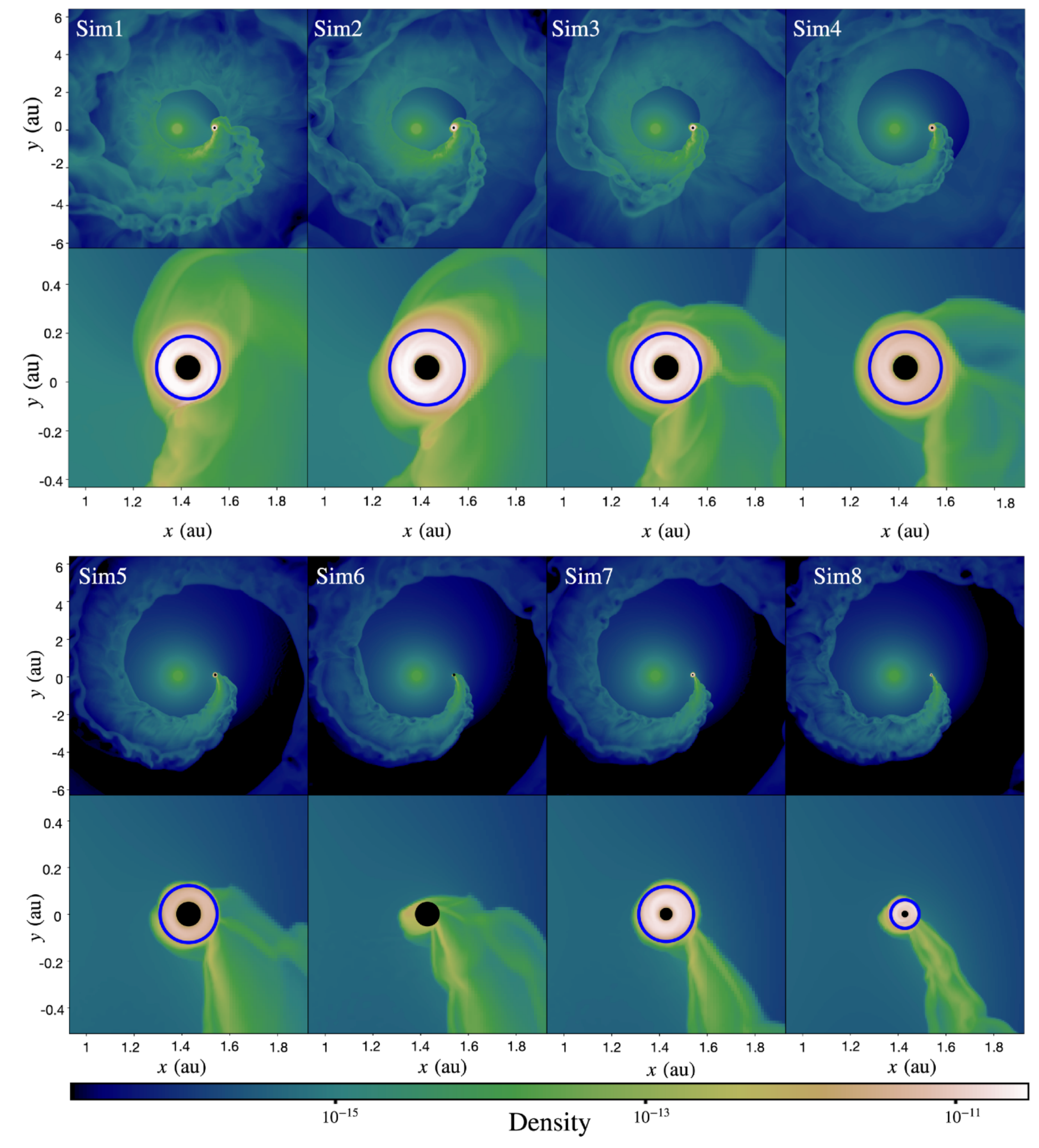}
  \caption{\label{fig:models}
    Density distributions obtained from the eight simulations, Sim1 through Sim8. For each simulation, the upper panel covers the entire simulation domain displaying the outflowing wind morphology in the orbital plane, while the lower panel is the zoomed-in version to clearly show the accretion flows around the WD. The simulation model is labeled at the top left corner of each upper panel. For each model, a snapshot at the last orbit is chosen. In the coordinate system where the center of mass of the binary system coincides with the coordinate center and the WD is on the $+x$-axis at $x=1.43$~au, the giant is located at $x=-0.57$~au. The outer rim of accretion disk is denoted by the blue circles. The black filled circles  in the lower panels indicate the accretion sink in the simulation.}
\end{figure*}

In the comparison of all simulated models, the Sim5 model exhibits a transitional morphology (see Fig.~\ref{fig:models}). The bow shock surrounding the accretion disk is well developed in the forward direction of the orbital motion of the WD in the slower Sim1--Sim4 models, while the material above the disk radius and beneath the bow shock is stripped out in the faster Sim5--Sim8 models. As the result, the accretion disks in the Sim5--Sim8 models are exposed to the winds and the accretion tails (or accretion columns) are revealed in the way that is demonstrated in the BHL theory \citep{BH44}. Notice that the morphological transition occurs with the {\it in-situ} intrinsic wind velocity being similar to the orbital velocity of the WD ($V_{\rm w}\sim V_{\rm WD}$).

\begin{figure*} 
  \plotone{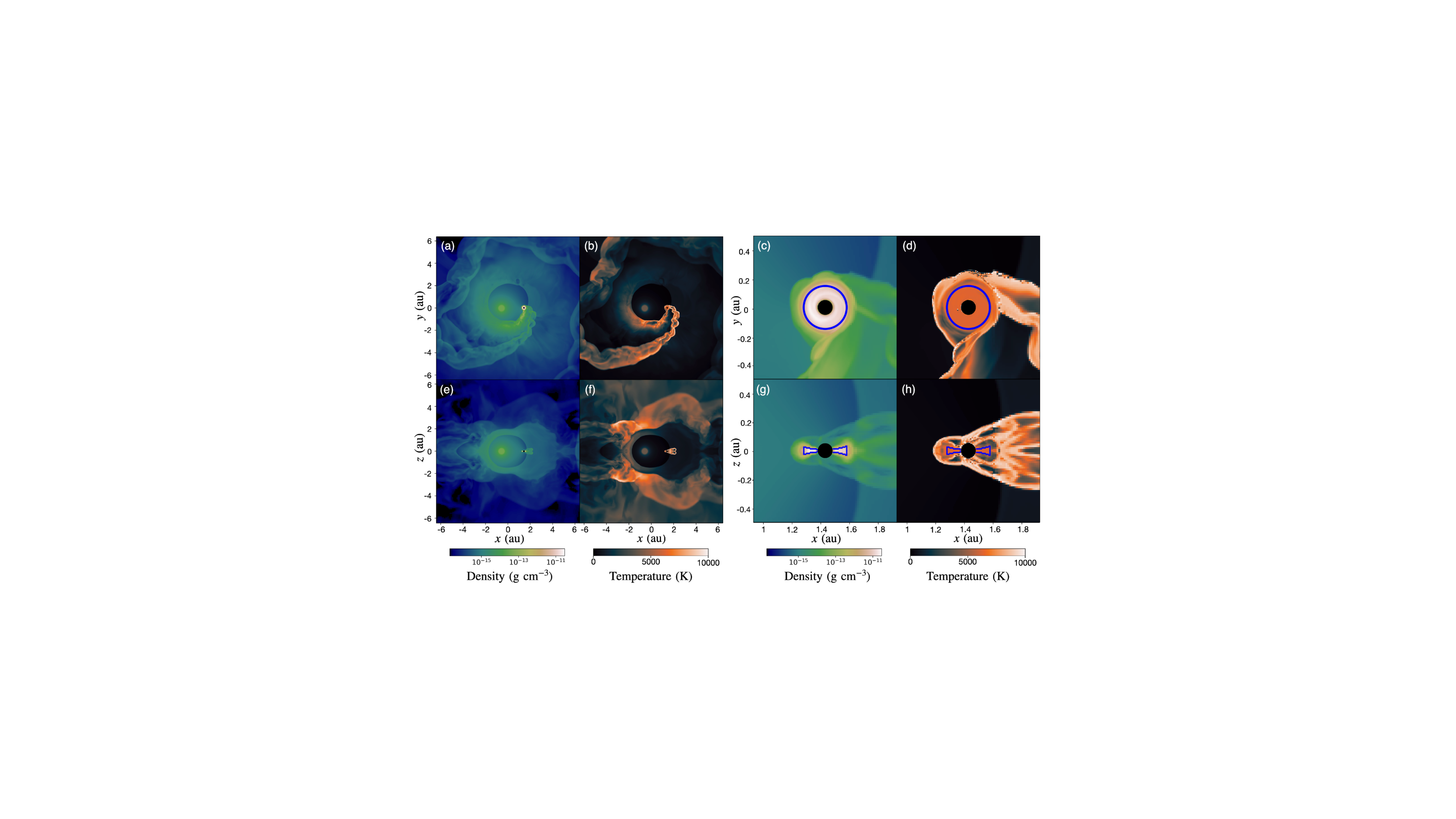}
  \caption{\label{fig:outflowing}
    Snapshots of the density and temperature distributions of the Sim2 model when the binary stars have completed their 4 orbital motions. Top panels from (a) to (d) exhibit the distributions in the equatorial $x$--$y$ plane, while bottom panels from (e) to (h) display the vertical structures in the $x$--$z$ plane. Panels (a), (b), (e) and (f) present the entire simulation domain including the spiral structure with its head at the WD, while panels (c), (d), (g) and (f) are the corresponding zoomed images targeting the accretion disk surrounding the WD. The outer boundary of the accretion disk, indicated by blue lines, is characterized by the radius of 0.16~au and the height of 0.03~au with a flared shape in the vertical direction. The snapshot images are centered at the center of mass of the binary star system.}
\end{figure*}

Figure~\ref{fig:outflowing} presents, as an example, the detailed structures of large to small scales of the Sim2 model at the time that the stars have completed their 4 orbits. The first and second columns of the figure exhibit the density and temperature distributions, respectively, viewed face-on (upper panels) and edge-on (lower panels). The third and fourth columns show the zoomed-in images of the same quantities near the WD. The binary stars are rotating in the counterclockwise direction around their center of mass at $(x,\,y,\,z) = (0,\,0,\,0)$. The giant star and the WD are currently located at $(x_{\rm G},\,y_{\rm G})=\rm(-0.57\,au,\,-0.01\,au)$ and at $(x_{\rm WD},\,y_{\rm WD})=\rm(1.42\,au,\,0.02\,au)$, respectively. 

It is clearly identified that a spiral structure attached to the WD coils around the central stars \citep[see also][]{theuns93, kim12b, huarte13, saladino18, saladino19a}. The inner and outer edges of the spiral structure are well defined by the narrow regions showing high temperature ($T>8000$~K; see Fig.~\ref{fig:outflowing}(b)). The outer edge is extremely turbulent and makes a thick wall filled with high temperature substructures. The inner edge is relatively smoothly distributed in both density and temperature. 
This inner edge is tightly wound around the giant star terminating at the meeting point with the outer edge of the spiral, $(x,\,y)\sim\rm(1.8\,au,\,0.2\,au)$, as seen in Figure~\ref{fig:outflowing}(a) and (c). Toward this position the temperature of the inner edge is no longer high (see Fig.~\ref{fig:outflowing}(b)). The region around the giant star confined by the inner spiral edge is relatively low in density.

Most of the wind material entering into the spiral structure eventually escapes from the binary potential, while some fraction of the material remains gravitationally captured by the WD, forming an accretion disk. The blue lines in the third and fourth columns of Figure~\ref{fig:outflowing} demonstrate the three dimensional morphology of the accretion disk (as defined in Section~\ref{sec:Formation}). Interestingly, the accretion disk is vertically not razor thin but is in a flare shape (Fig.~\ref{fig:outflowing}(g)--(h)).

\subsection{Formation of accretion disk}\label{sec:Formation}

The formation criterion of an accretion disk adopted in this work is the presence of circularized accretion flows around the WD that last longer than an orbital period. If formed, the shape and mass of an accretion disk are computed as following. 

\begin{itemize}
\item {\bf Disk height $H_{\rm disk}$.} We first draw ten density profiles along the vertical lines passing the points of intersection with the equatorial plane at an arbitrary radius $R$ from the WD, starting from $r_{\rm sink}$ and increasing with an intervals of $\Delta R$. The gas density decreases following an exponential function up to a certain height, and beyond this height the density abruptly drops. We define the scale height $H(R)$ at a radius $R$ as the average of the ten values for these heights characterizing the exponential density declines. The disk height $H(R_{\rm disk})$, or simply $H_{\rm disk}$, is the scale height at the radius of the disk, $R=R_{\rm disk}$, defined as below.

\item {\bf Disk mass $M_{\rm disk}$.} We estimate the mass $M(R)$ within the volume of the flare-shaped disk, bounded by a cylinder with an arbitrary radius $R$ and the scale height $H(R)$, with its inner boundary at the sphere of radius $r_{\rm sink}$. We repeat this mass estimation by gradually raising up the radius of the outer boundary by $\Delta R$ until the mass increment is negligible (less than 1\%), i.e., $M(R+\Delta R) - M(R) < 0.01 M(R+\Delta R)$, and the final value is defined as the disk mass $M_{\rm disk}$, or $M (R_{\rm disk})$.

\item {\bf Disk radius $R_{\rm disk}$} is correspondingly defined as the radius at which the above mass condition is satisfied.
\end{itemize}

\begin{figure} 
  \plotone{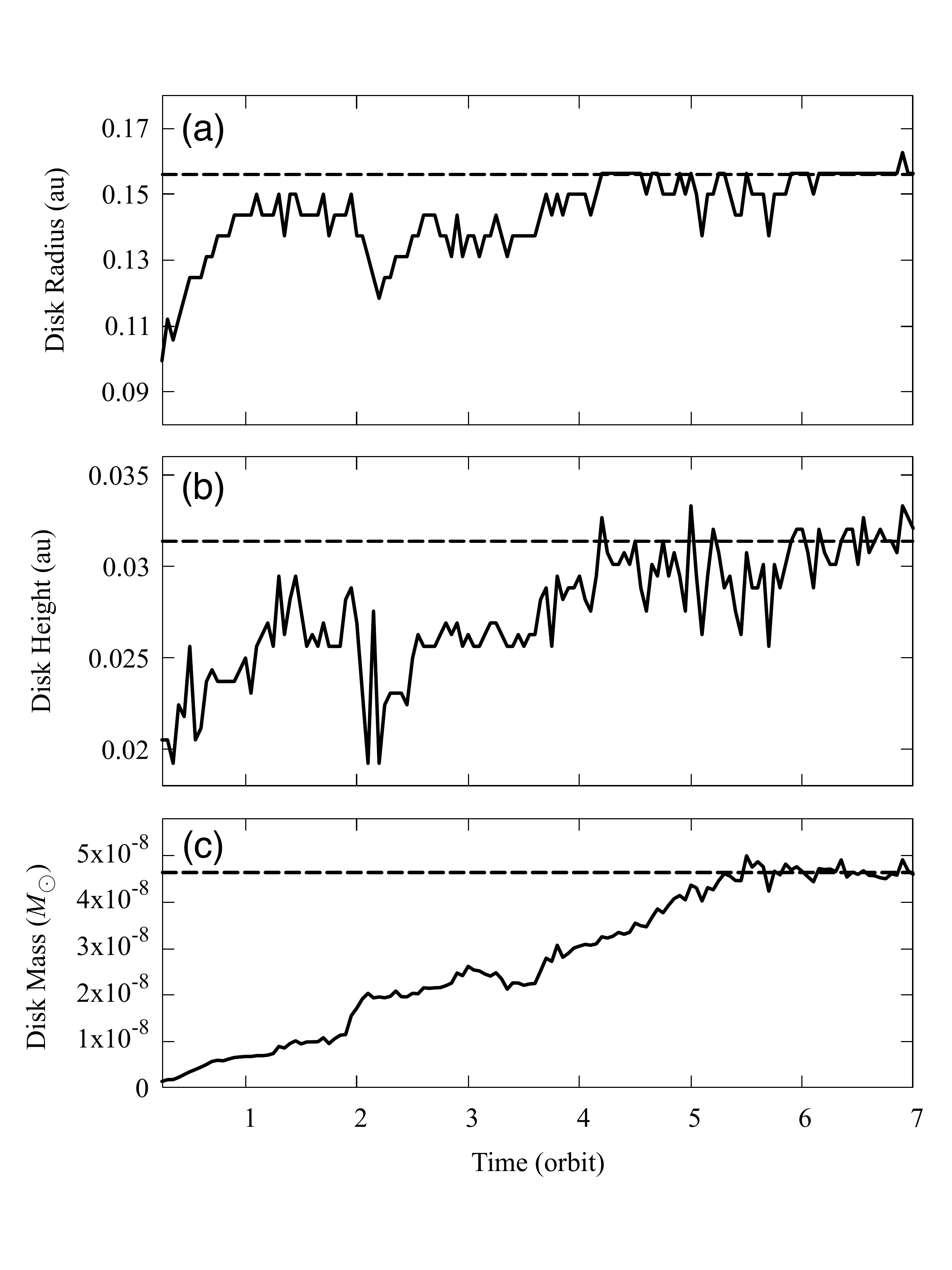}
  \caption{\label{fig:m2size}
    Temporal evolution of the (a) radius, (b) height, and (c) mass of the accretion disk for the Sim2 model. The horizontal dashed lines denote the final stable values, obtained by averaging the quantities during the last orbit. The disk radius and height are nearly stabilized at $t\sim4$ orbit, while the disk mass reaches the final value at $t\sim5.4$ orbit.}
\end{figure}

We estimate these three quantities at each simulation time step. The values show some fluctuations in the early orbits and become stabilized in the later orbits. In Figure~\ref{fig:m2size}, $R_{\rm disk}$, $H_{\rm disk}$, and $M_{\rm disk}$ of the, for example, Sim2 model are displayed as a function of time. The disk radius $R_{\rm disk}$ increases slightly beyond the noise level of variation until eventually reaching a nearly constant value of $\sim0.15$~au at $t>4$ orbit. The disk height $H_{\rm disk}$ shows a $\sim40\%$ increase in the early evolution ($t<4$ orbit) and attains a near-constant value of $\sim0.03$~au. 
The disk mass reaches up to $M_{\rm disk} \sim 4.6\times10^{-8}\,\Msun$ at $t\sim5.4$ orbit and stops increasing afterwards. The final disk mass corresponds to $\sim3\%$ of the mass expelled from the giant star for 14 years (7 orbits). We note that the growth of disk mass within the nearly stationary volume of the disk at $t=4$--5.4 orbit enhances the density in the disk (see the change in the minimum of density profile along time in Section~\ref{sec:totAsym}).
The mean values of these quantities over the last orbit in the simulations are listed in Table~\ref{tab:totmodel}.

The disks are always formed regardless of the initial parameter sets, if the simulation mesh resolution and the accretion sink radius are properly assigned.
At the same spatial conditions with the Sim1--Sim5 models, the Sim6 model fails to create an accretion disk (see Fig.~\ref{fig:models}). It turns out to be a numerical artifact, as the Sim7 model with a reduced sink radius $r_{\rm sink}(=r_{\rm soft, WD})$ by half successfully forms an accretion disk with the radius of 0.1~au. From these experiments, we note that the sink radius greater than $\sim25\%$ of the (potential) disk radius causes an artificial destruction of the disk. Here, we did not change the resolution of simulation grids as it satisfies the minimum requirements for resolving $r_{\rm sink}$ (i.e., 4 cells within $r_{\rm sink}$).
With the Sim8 model, we also test the disk formation in the environment with the adjacent wind velocity faster than the orbital velocity of the WD. In this model, by reducing $r_{\rm sink}$ and $r_{\rm soft, WD}$ to 0.0125 au, we successfully produce the accretion disk with its radius of 0.058~au.

\begin{figure} 
  \plotone{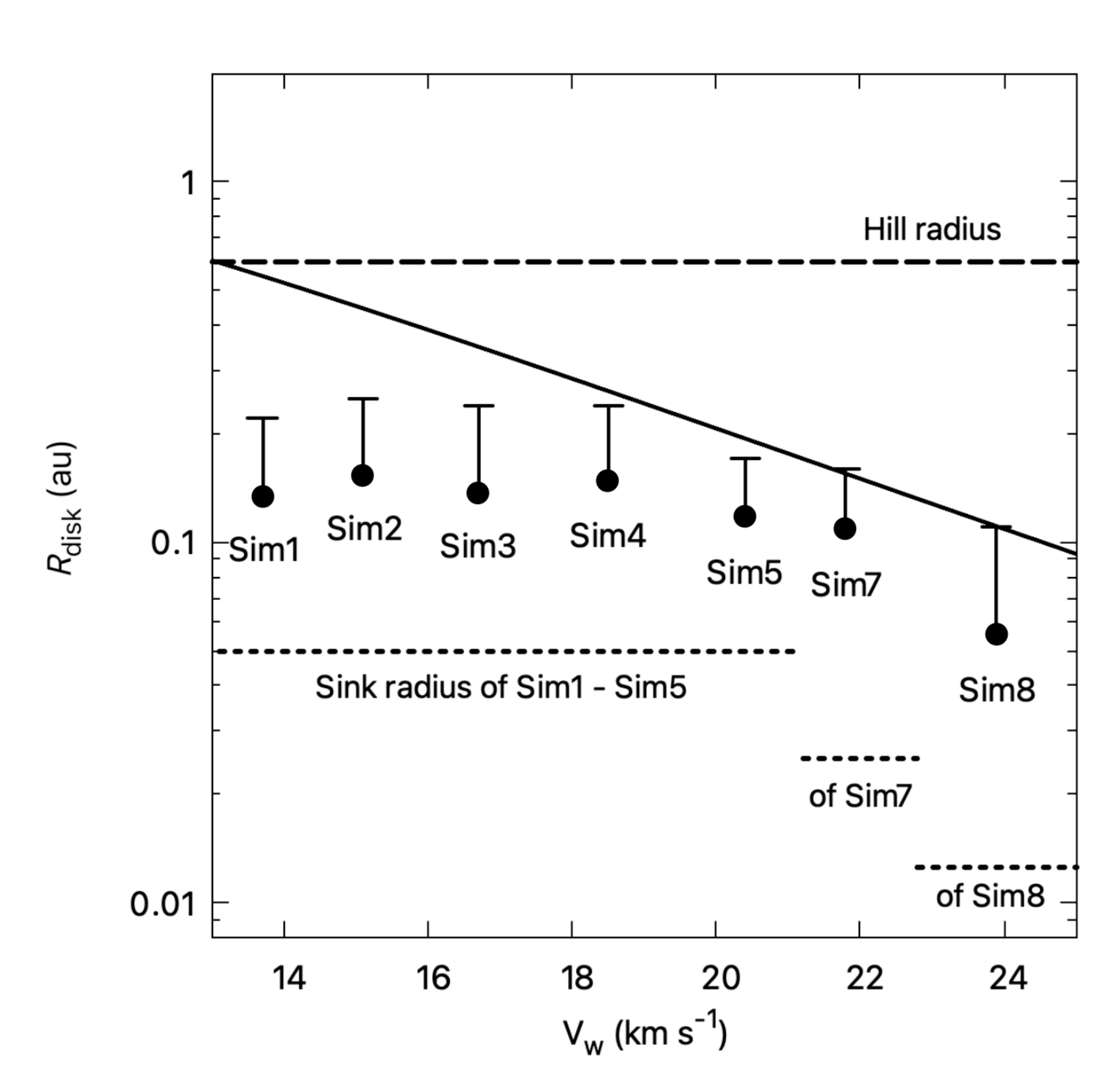}
  \caption{\label{fig:CompHE}
    Radii of accretion disks (filled circle) and stand-off distances of the bow shocks (horizontal bar) obtained from the simulations. The solid line shows  the accretion column impact parameter $b$ proposed as an upper limit for the disk radius by \citet{huarte13}. For comparison, the Hill radius of the simulated binary system is marked by the horizontal long-dashed line and the sink radii for Sim1--Sim5, Sim7, and Sim8 are shown by the three horizontal short-dashed lines.}
\end{figure}

As noted in Table~\ref{tab:totmodel}, the representative disk radius (i.e., the average of time oscillation of disk radius over the last simulation orbit) does not change with the wind velocity in the Sim1--Sim4 models but starts to decrease at the faster winds in the Sim5--Sim8 models.
The disk radius $R_{\rm disk}$, tabulated in Table~\ref{tab:totmodel} and indicated by filled circles in Figure~\ref{fig:CompHE}, is smaller than the accretion line impact parameter $b$ (black solid line in Fig.~\ref{fig:CompHE}) derived by \citet{huarte13} for the BHL flow toward the retarded position of the WD:
\begin{equation}\label{eq:he}
  b = 2\,q^2\,(1+q)^3 \left( \frac{1}{a} \right)^{5/2}
  \left( \frac{1}{a} + \frac{1}{a_w} \right)^{-7/2},
\end{equation}
where $q=M_{\rm WD}/M_{\rm G}$ is the stellar mass ratio, $a$ is the binary separation, and $a_w=GM_{\rm G}/((1+q)V_{\rm w}^2)$ is a specific length scale indicating the binary separation when the wind velocity $V_{\rm w}$ equals to the orbital velocity of the WD about the center of mass of binary stars, i.e., $V_{\rm w}=V_{\rm WD}(a=a_w)$. 
The disk radii of the Sim1--Sim4 models are almost the same and significantly smaller than the $b$ parameter, while the disk radii of the Sim5--Sim8 models decrease following the decreasing trend of the $b$ line along the wind velocity. 
On the other hand, the bar above the filled circle symbol in Figure~\ref{fig:CompHE} presents the distance to the outermost rotating component of material around the WD, measured along the line toward the giant star; in most of cases, it represents the stand-off distance of the bow shock. Figure~\ref{fig:CompHE} indicates that the b parameter actually limits this distance, which, by definition, is larger than the disk radius.
For reference, the Hill radius and the sink radius of each model are also drawn in Figure~\ref{fig:CompHE} in order to show the definitely forbidden area for the disk radius.

\subsection{Accretion rate}\label{sec:AccRate}

The mass accretion rate $\dot M_{\rm acc}$ is measured by integrating the mass entering into the sink sphere with its energy condition that forbids its escape from the sink. 
For comparison, we also measure the mass transfer rate $\dot M_{\rm inflow}$ by integrating mass fluxes inflowing through the surface of the flared disk (defined in Section~\ref{sec:Formation}), and the disk mass $M_{\rm disk}$ is independently measured in Section~\ref{sec:Formation} by integrating the density distribution within the flared disk excluding the sink sphere.
In our fiducial Sim2 model, the averaged rates over the entire simulation time are $\dot M_{\rm inflow} \sim 4.8\times10^{-8}\ \Mspy$ and $\dot M_{\rm acc} \sim 2.3\times10^{-8}\ \Mspy$, implying that the inflowing matter to the disk is twice the matter swallowed by the WD. Because $\dot M_{\rm inﬂow}$ exceeds $\dot M_{\rm acc}$, the inflowing matter is gradually accumulated within the disk and therefore the disk mass increases along the evolution, which is consistent with Figure~\ref{fig:m2size}(c). 

\begin{figure} 
  \plotone{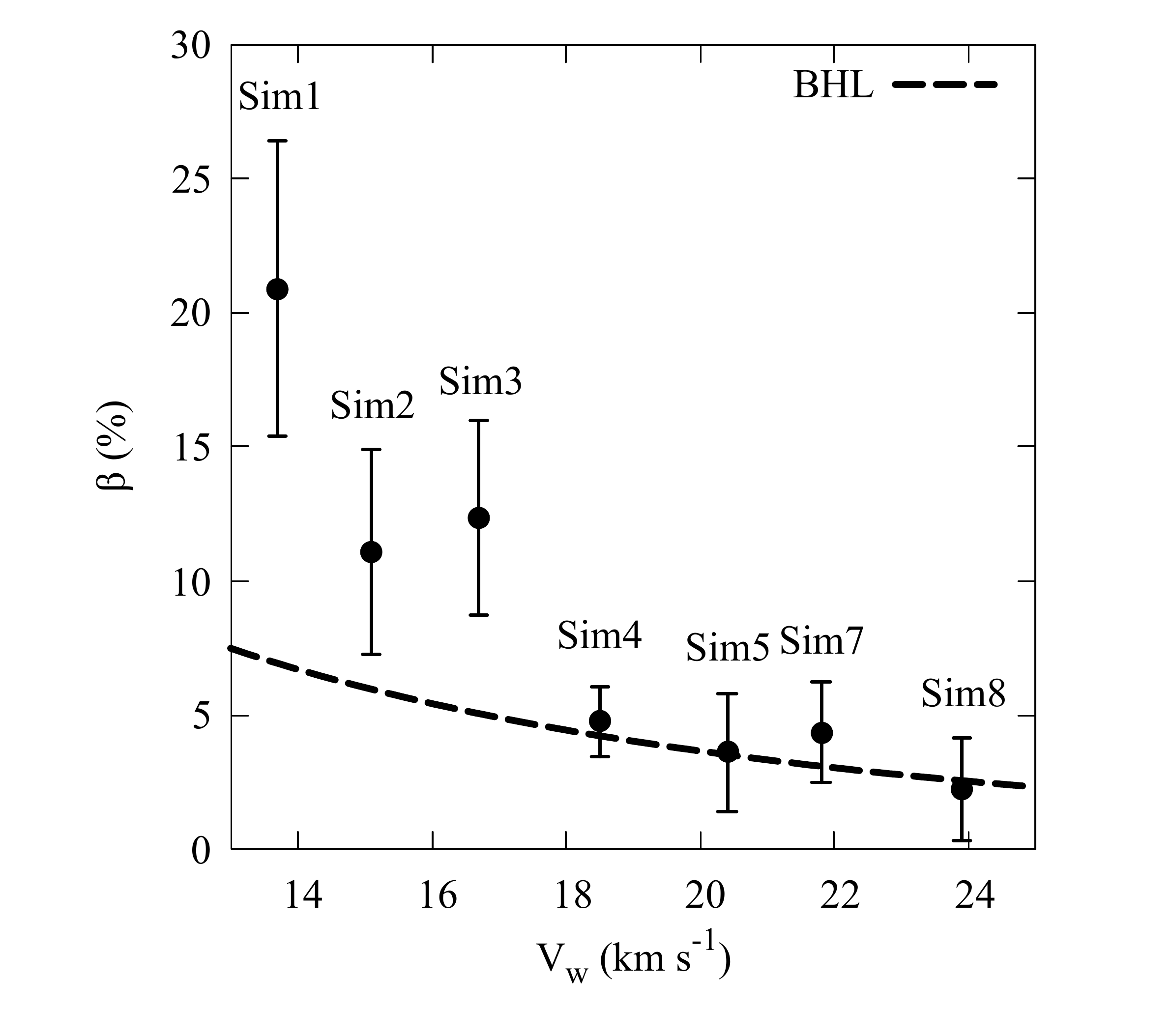}
  \caption{\label{fig:Accrate}
    Accretion efficiency $\beta$ as a function of the wind velocity $V_{\rm w}$. The accretion efficiency is defined as the ratio of the rate of mass entering the accretion sink and the mass-loss rate of the giant star. The filled circles are our model results and the dashed curve shows the efficiency $\beta_{\rm BHL}$ from the Bondi-Hoyle-Lyttleton theory (see the text).}
\end{figure}

Figure~\ref{fig:Accrate} presents the time-averaged value of mass accretion efficiency $\beta$, which is defined as the mass accretion rate into the sink of WD divided by the mass loss rate from the giant star. To avoid spikes in the measurements of accretion rates, the averages and errors are estimated in the median base.
The accretion efficiency is higher than 10\% in the Sim1--Sim3 models, and it rapidly decreases along the wind velocity up to $\sim18\ \kmps$. The accretion efficiency remains at a level of a few percent in the Sim4--Sim8 models. These measurements in our simulation models are compared with the accretion efficiency under the BHL assumption \citep{boffin88}:
\begin{equation}
  \beta_{\rm BHL} = \frac{\alpha_{\rm BHL}}{2} \left( \frac{q}{1+q} \right)^2
  \left( \frac{V_{\rm rel}}{V_{\rm w}} \right)^4 
  \left[ 1 + \left( \frac{V_{\rm rel}}{V_{\rm w}} \right)^2 \right]^{-3/2},
\end{equation}
where $V_{\rm rel}=(1+q)\,V_{\rm WD}$ is the orbital velocity of the WD in the rest frame of the giant star, and the efficiency coefficient $\alpha_{\rm BHL}=1$ is adopted. It is found that the accretion efficiencies in the slow wind models (Sim1--Sim3) are considerably larger than $\beta_{\rm BHL}$, while those in the faster wind models (Sim4--Sim8) are similar to the BHL prediction. 

\citet{espey08} pointed out the poor understanding of the mass loss processes in cool giants quoting the wind terminal velocities $<100\ \kmps$ \citep[also e.g.,][]{dupree87}. In the case of S-type symbiotic stars with the wind velocity exceeding $20\ \kmps$, our simulations imply that the accretion efficiency will be a few percent, comparable with that of the BHL accretion. However, a higher accretion efficiency of $\sim 10$ percent or more is expected for S-type symbiotic systems with a lower wind velocity than the orbital velocity of WD. The orbital parameters for most D-type symbiotic stars are only poorly known so that a reliable estimate of the accretion efficiency is not obvious based on the current work 
\citep[e.g.,][]{schmid02, matthews06, hinkle13}.

\subsection{Disk spirals}\label{sec:m2spiral}

Of particular interest is the presence of two inflowing spirals within the accretion disk (see e.g., Fig.~\ref{fig:outflowing}(c)).
In cataclysmic variables, the disk spirals are excited by tidal interaction, as being confirmed by many observations through the indirect imaging technique such as Doppler tomography and inviscid hydrodynamic calculations at several non-adiabatic equations of state \citep[][and references therein]{matsuda2000}. 
In symbiotic stars, however, an observational confirmation of the disk spirals was not made and the theoretical approach to the disk spirals was only performed by \citet{bisikalo1997}. 
In this work, we stress that the disk spirals can be formed even in symbiotic stars through the 3D wind accretion calculation for the first time.

\begin{figure*} 
  \plotone{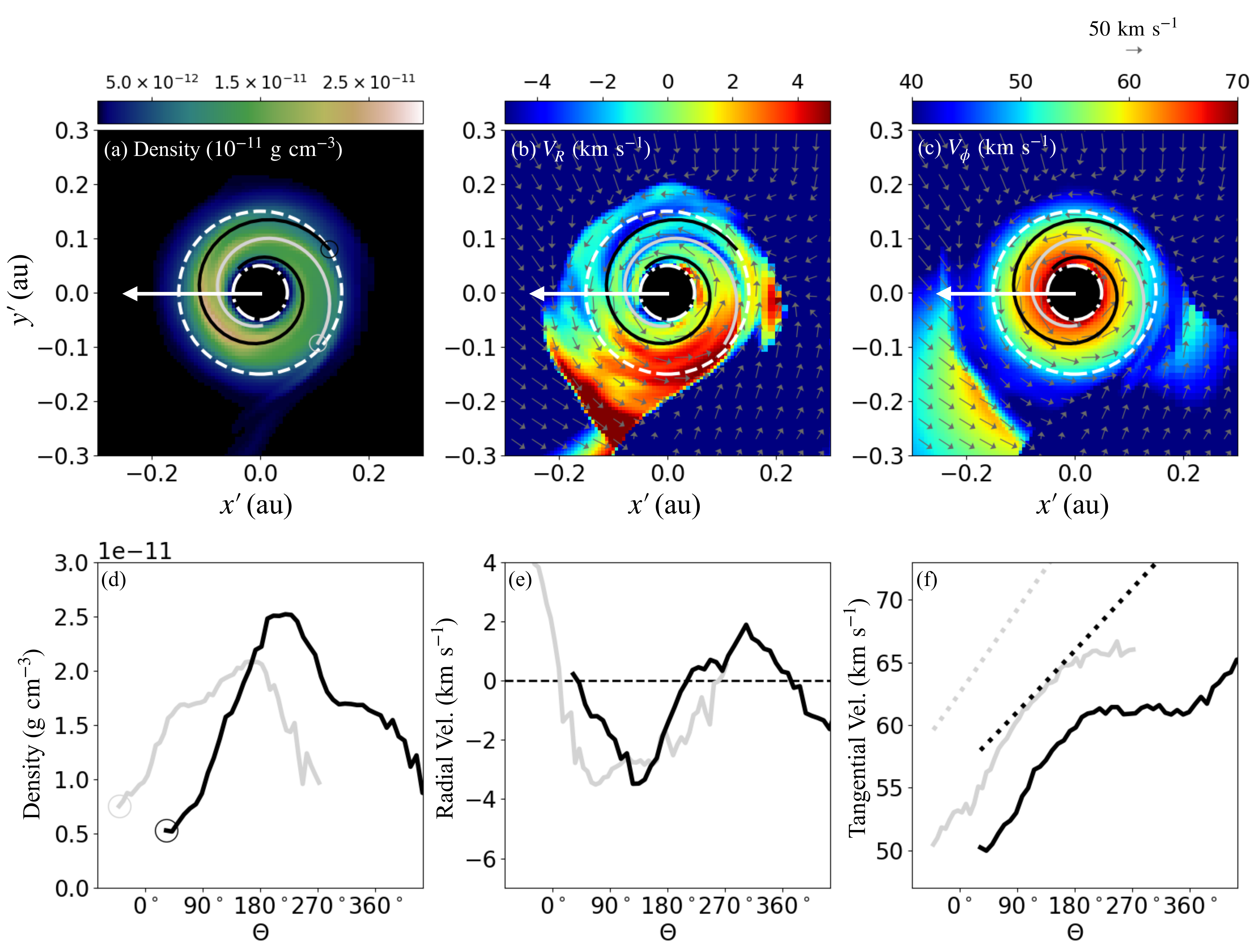}
  \caption{\label{fig:a2v9Disk}
    Maps of (a) density, (b) radial velocity $V_{R}$, and (c) rotational velocity $V_{\phi}$ around the accretion disk of the Sim2 model at $t=4$ orbit. The long white arrow points the direction toward the giant star located at $(x^\prime,\,y^\prime)=\rm(-2\,au,\,0\,au)$, and the orbital motion of the WD is along the $+y^\prime$ direction. The velocities are measured in the rest frame of the WD. The white dashed and dash-dotted lines denote the outer rim of the disk and the accretion sink, respectively. 
    By tracing the local density maxima and extracting the best fit spiral in the form of a logarithmic spiral function, we find the two spiral structures, which are designated as the {\it front spiral} (black line) and {\it rear spiral} (gray line). Panels (d) to (f) show the corresponding physical quantities along these two spirals as a function of angular position $\Theta$ (black for {\it front spiral}; gray for {\it rear spiral}). The horizontal dashed line in Panel (e) corresponds to $V_R=0$. The dotted lines in Panel (f) shows the Keplerian velocities at the positions along the two spirals.}
\end{figure*}

We present, for example, the Sim2 model in Figure~\ref{fig:a2v9Disk} (a) the maps of the density in a linear color scale, (b) the radial velocity $V_R$, and (c) the rotational velocity $V_\phi$ of the matter. The magnitudes of directional velocities in the (b)--(c) maps and the fluid velocity vectors denoted by the gray arrows are all recalculated in the rest frame of the WD at the current time $t=4$ orbit. 
One spiral entering into the disk radius at $\Theta\sim30\arcdeg$, i.e., from the forward direction of the orbital motion of the WD ({\it front spiral}, hereafter), is winding the WD for more than one lap before reaching the sink radius (see black solid curve in Fig.~\ref{fig:a2v9Disk}(a)--(c)). The other spiral presents a shorter trajectory starting from $R=R_{\rm disk}$ at $\Theta\sim-40\arcdeg$ ({\it rear spiral}; see gray solid curve).

In order to measure the physical quantities along the spirals, we first define the shapes of the disk spirals in a functional form of logarithmic spiral: $R/R_{\rm disk}=\exp[-(\Theta-\Theta_a)/\Theta_b]$, where the $\Theta_a$ parameter satisfies $\Theta_a=\Theta(R=R_{\rm disk})$ and the polar slope of the spiral is a constant $-\Theta_b^{-1}$. After unrolling the density map of the accretion disk in the polar coordinates, we find the positions of local density maxima along the radius axis within a 5-pixel margin from the test function of a logarithmic spiral. The $\chi^2$ goodness of fit test provides the coeﬃcients: $(\Theta_a,\,\Theta_b)=(33\arcdeg\pm4\arcdeg,\,500\arcdeg\pm10\arcdeg)$ for the {\it front spiral} and $(\Theta_a,\,\Theta_b)=(-41\arcdeg\pm3\arcdeg,\,332\arcdeg\pm8\arcdeg)$ for the {\it rear spiral}, where the radial distance $R$ from the WD is defined from the disk radius $R_{\rm disk}=0.15$~au to the sink radius $r_{\rm sink}=0.05$~au.

Figure~\ref{fig:a2v9Disk}(d) shows that the densities along these spirals are $\sim 5\times10^{-12}\rm\ g\,cm^{-3}$ at $R\sim R_{\rm disk}$ (marked by open circles), as being similar to each other. The density along the {\it rear spiral} increases up to $\sim 2\times10^{-11}\rm\ g\,cm^{-3}$ at $\Theta \sim 180\arcdeg$ and then steeply decreases until the spiral reaches $R=r_{\rm sink}$. On the other hand, the density along the {\it front spiral} increases up to $\sim 2.5 \times 10^{-11}\rm\ g\,cm^{-3}$ at $\Theta \sim 220\arcdeg$--\,250\arcdeg\ and then decreases until the spiral reaches the sink radius. 

It is noticeable that the density along the {\it front spiral} ceases the decreasing trend at $\Theta \sim 300\arcdeg$--\,360\arcdeg\ (see Fig.~\ref{fig:a2v9Disk}(d)). Interestingly, in the similar range of $\Theta$, the radial velocity of matter along the {\it front spiral} becomes positive at $\Theta \sim 220\arcdeg$--\,370\arcdeg\ (see Fig.~\ref{fig:a2v9Disk}(e)) and the rotational velocity ceases its increasing trend by staying at $V_\phi \sim 60\ \kmps$ at $\Theta = 200\arcdeg$--\,360\arcdeg\ (see Fig.~\ref{fig:a2v9Disk}(f)). 
Our best interpretation for these abnormal behaviors of the {\it front spiral} is that the flows entering the disk from the front side, with respect to the orbital motion of the WD, are not well circularized but are slightly overshooted toward the rear side ($\Theta \sim 270\arcdeg$), perhaps due to the complex process during the convergence of the streamlines (see Section~\ref{Streamlines}). The overshooted matter raises the neighboring density beyond the ordinary value, and then the matter is diverted onto the outer disk region until falling back to its ordinary path beyond $\Theta \sim 360\arcdeg$.
In contrast, the {\it rear spiral} does not show such a drifting event and a clear spiral-in feature of it is implied by the gray-colored profiles in Figure~\ref{fig:a2v9Disk}(d)--(f). 

\subsection{Accretion Streamlines}\label{Streamlines}

\begin{figure*} 
  \plotone{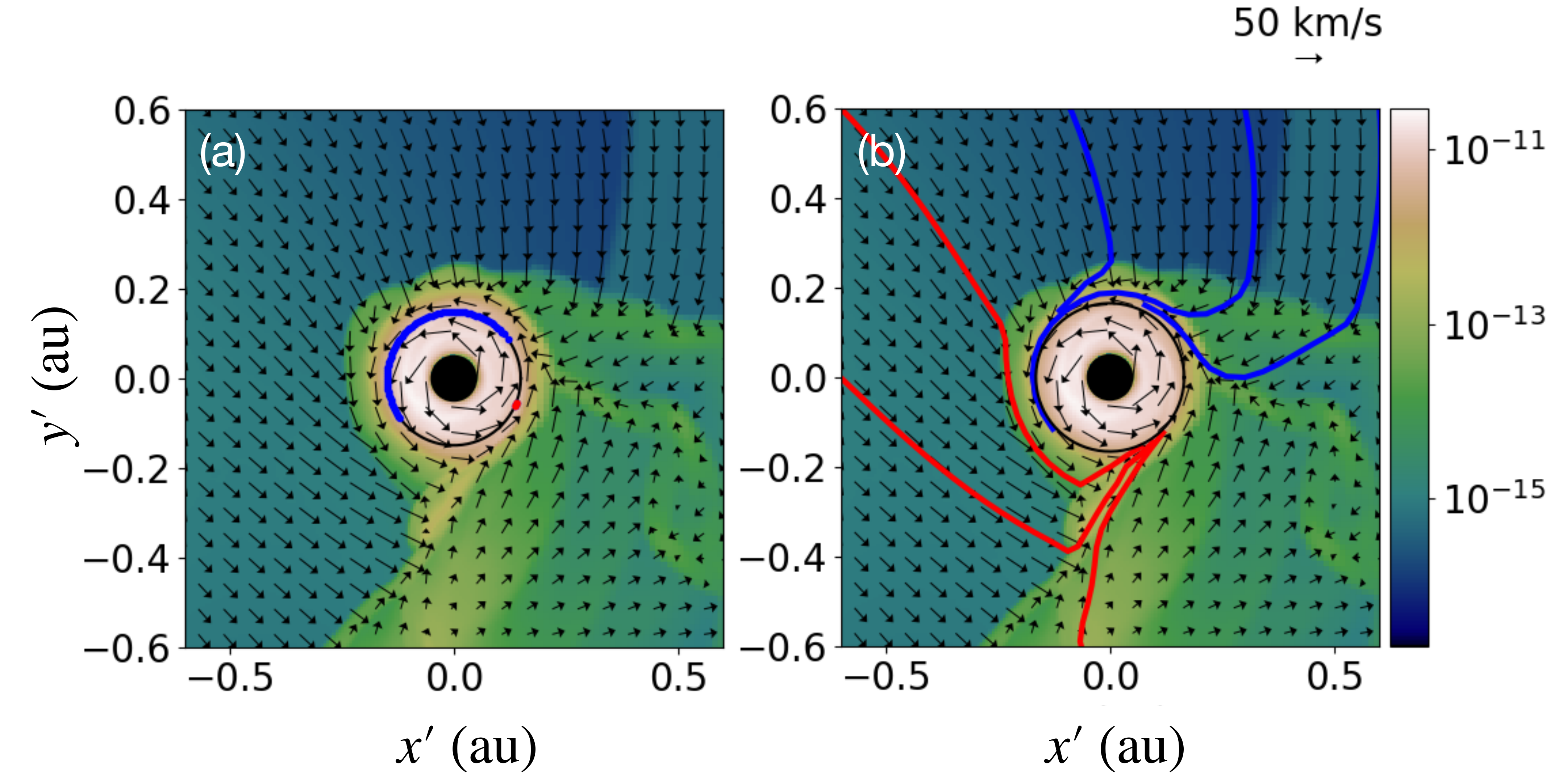}
  \caption{\label{fig:streamlines}
    In Panel (a), the two arc segments at $R=R_{\rm disk}$, through which matter flows in, are shown by blue and red colors. Panel (b) shows 6 representative streamlines that extend to the outer rim of the disk. The blue streamlines pass through the bow shock to join the disk at the blue arc segment indicated in (a). The red streamlines circumvent the bow shock to reach the small red arc segment shown in (a). The background map and arrows present the density distribution in units of {g\,cm$^{-3}$} and the velocity vectors of the flows in the equatorial plane of Sim2 model at $t=4$ orbit. Here, the velocity is measured in the rest frame of the WD.}
\end{figure*}

A close inspection of streamlines reveals that there are two major streams acting as the origin of disk material. For a streamline analysis, we trace the streamlines starting from 400 random points far from the disk. When the flows arrive at the disk radius with negative (i.e., incoming) radial velocities, we record those entering spots at the disk radius $R_{\rm disk}$ in Figure~\ref{fig:streamlines}(a). The arrows in this figure show the magnitudes and directions of the fluid velocities in the rest frame of the WD. 

We find that the flows do not enter the disk through the entire outer rim of the disk but the entering spots are instead limited to the two arc segments  marked by the blue and red colors in Figure~\ref{fig:streamlines}(a). The angular extent of the blue arc segment ranges from 36\arcdeg\ to 216\arcdeg\ spanning a half circle, whereas the red arc segment has a small angular extent of $\Delta\Theta \sim 3\arcdeg$ limited to the range between 336\arcdeg\ and 339\arcdeg.

Figure~\ref{fig:streamlines}(b) shows six representative accretion stream lines that arrive at the outer rim of the disk. It is apparent that the three red stream lines converge at the red arc segment shown in Figure~\ref{fig:streamlines}(a). The other three blue streamlines represent the wind components originating directly from the giant star, which subsequently get deflected  as they pass through the bow shock. The entering spots of this blue group of flows are spread over $\Delta\Theta \sim 180\arcdeg$, as marked by the blue arc segment in Figure~\ref{fig:streamlines}(a). 
The inflows from the rear side amount to $\sim 50\%$ of the total incoming mass, which means that a similar amount mass enters the disk from the front side through the bow shock. 

The two groups of inflows appear to be assimilated into the two spiral features after entering the disk. The {\it rear spiral} is likely composed of the inflowing matter entering through the small arc segment, marked by red spots in Figire~\ref{fig:streamlines}(a), whose angular position is similar to $\Theta_a$ of the {\it rear spiral}. The inflows from the front side through the large arc segment (blue spots) would compose the {\it front spiral}, and the incessant injection of matter throughout a large fraction of the spiral would be responsible for the abnormal behavior of the {\it front spiral} described in the previous subsection.

\section{Observational Ramifications for AG~Dra}\label{sec:obs}

One important goal of this work is to obtain the matter distribution and the kinematics of the accretion flow, which will be used as the input parameters for line profile analyses of Raman-scattered \ion{O}{6} features in the future. In particular, our Sim2 simulation is carried out with a view to finding the physical properties of the accretion disk in the S-type symbiotic star AG~Dra in the quiescent phase.

\subsection{Asymmetric disk distribution}\label{sec:totAsym}

\begin{figure*} 
  \plotone{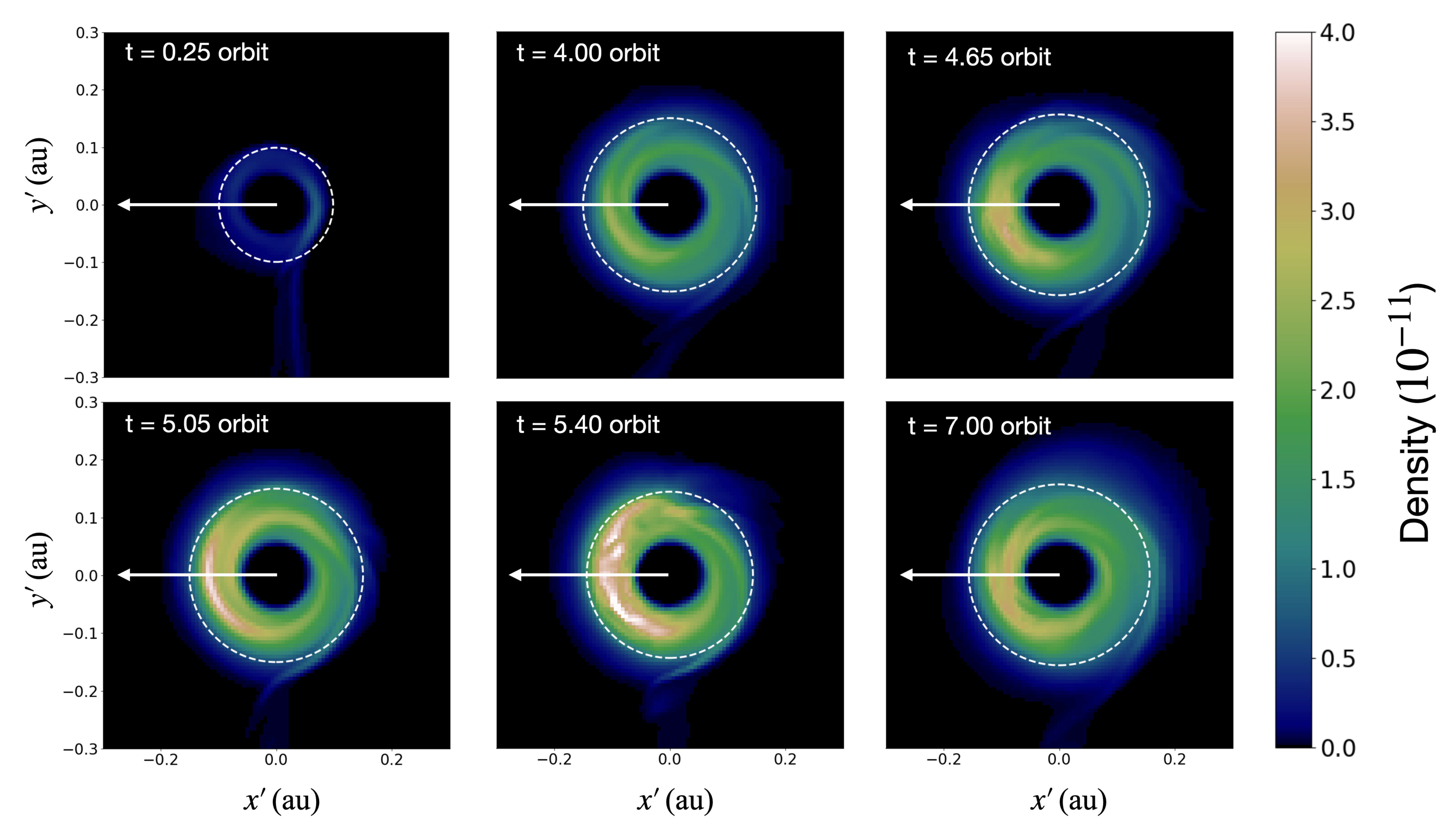}
  \caption{\label{fig:v9Evol}
    Snapshots of the density distribution in the equatorial plane of disk of Sim2 model. The time in units of the binary orbital period is denoted at the top left of each panel. The white dashed circle in each panel shows the measured disk radius; 0.10, 0.15, 0.16, 0.15, 0.14, and 0.16~au at $t=0.25$, 4.00, 4.65, 5.05, 5.40, and 7.00 orbits, respectively. The long white arrows indicate the direction toward the giant component.}
\end{figure*}

One of the notable features in disk shape is the asymmetric density distribution. Figure~\ref{fig:v9Evol} presents the time evolution of the density maps of the Sim2 disk model in the equatorial plane. We find that the overall density distribution in the disk is conspicuously asymmetric and the highest density is always achieved near the direction toward the giant star. In addition, the entering spots of the accretion streamlines do not show significant changes in the azimuthal angular coordinate, $\Theta$, measured with respect to the direction toward the giant star as $\Theta=180\arcdeg$. The two spirals repeat merging (at $\sim 4.65$ orbit and $\sim 5.40$ orbit) and splitting (at $\sim 4.00$ orbit and $\sim 5.05$ orbit) along their evolution, which shows a tendency of the highest degree of asymmetry when they merge.

\begin{figure*} 
  \plotone{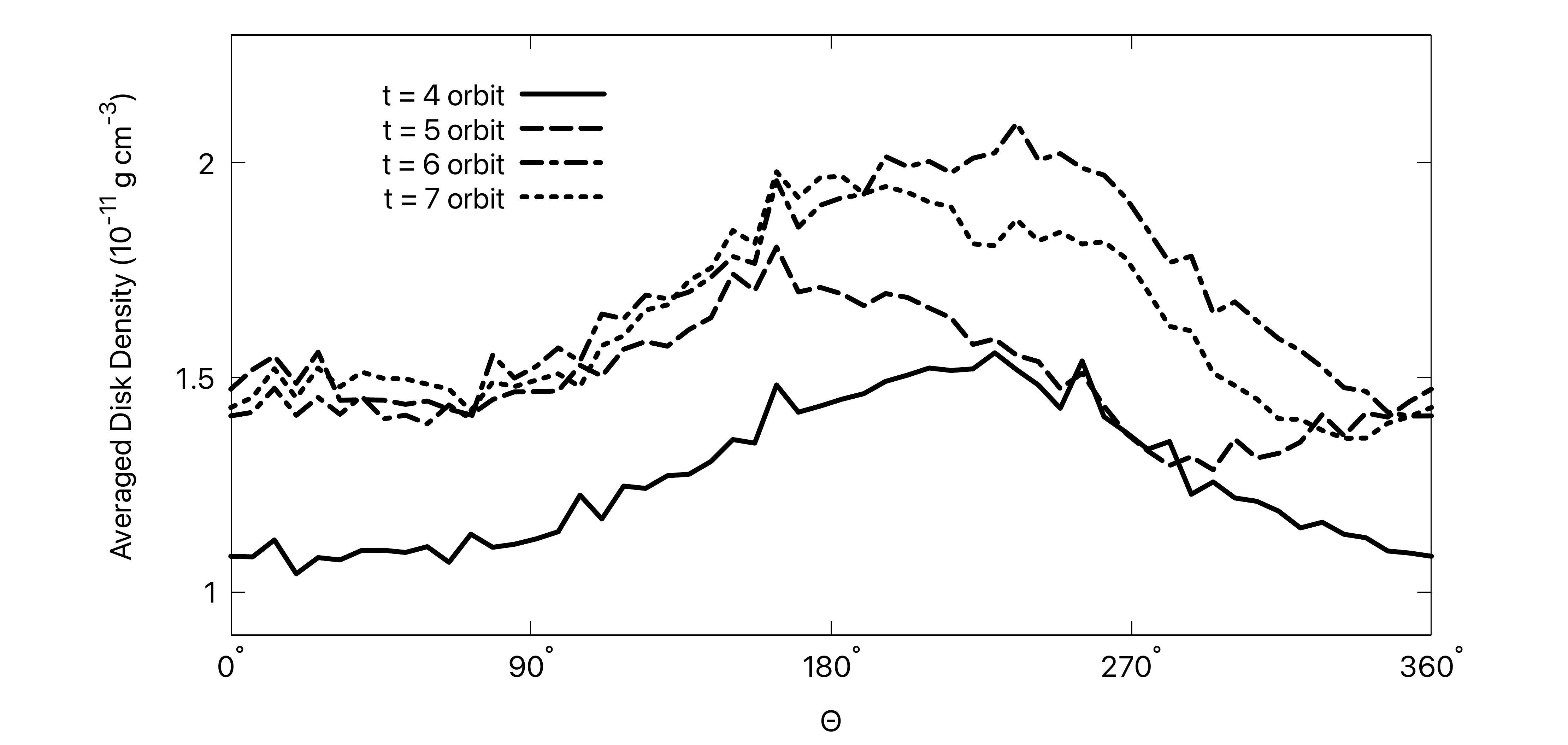}
  \caption{\label{fig:m2asymmetry}
    Density profiles in the equatorial plane averaged over the radius ranging from $r_{\rm sink}$ to $R_{\rm disk}$ from the selected snapshots at $t=4$, 5, 6, and 7 orbits for the Sim2 model. The horizontal axis shows the azimuthal angle $\Theta$, where $\Theta=180\arcdeg$ corresponds to the direction toward the giant star. The angular position $\Theta(\rho_{\rm max})$ of the density peak is found in the range of 170\arcdeg--\,230\arcdeg. The magnitude of density of the disk increases in the period between 4 and 5 orbit, and then remains unchanged afterwards.}
\end{figure*}

Figure~\ref{fig:m2asymmetry} shows the density profiles averaged over radius in the range $r_{\rm sink}<R<R_{\rm disk}$  for the Sim2 model. The density profiles are overall single peaked at $\Theta\sim218\arcdeg$, 175\arcdeg, 229\arcdeg, and 189\arcdeg\ at $t=4$, 5, 6, and 7 orbit, respectively. As we analyzed in Section~\ref{sec:m2spiral}, such density enhancement at the direction toward the giant star, or on the slightly rare side with respect to the orbital motion, is probably originated from the overshooting of the accretion flows composing the {\it front spiral}.
The maximum densities are $\sim50\%$ higher than the minima in the opposite sides of the disk. It implies that, if the emissivity depends only on the squares of density, the emissivity ratio between the maximum and minimum in a disk exceeds $\sim2$, which is consistent with the best-fitting result for the accretion disk of AG Dra hypothesized by \citet{lee19}.

\begin{figure} 
  \epsscale{0.75}
  \plotone{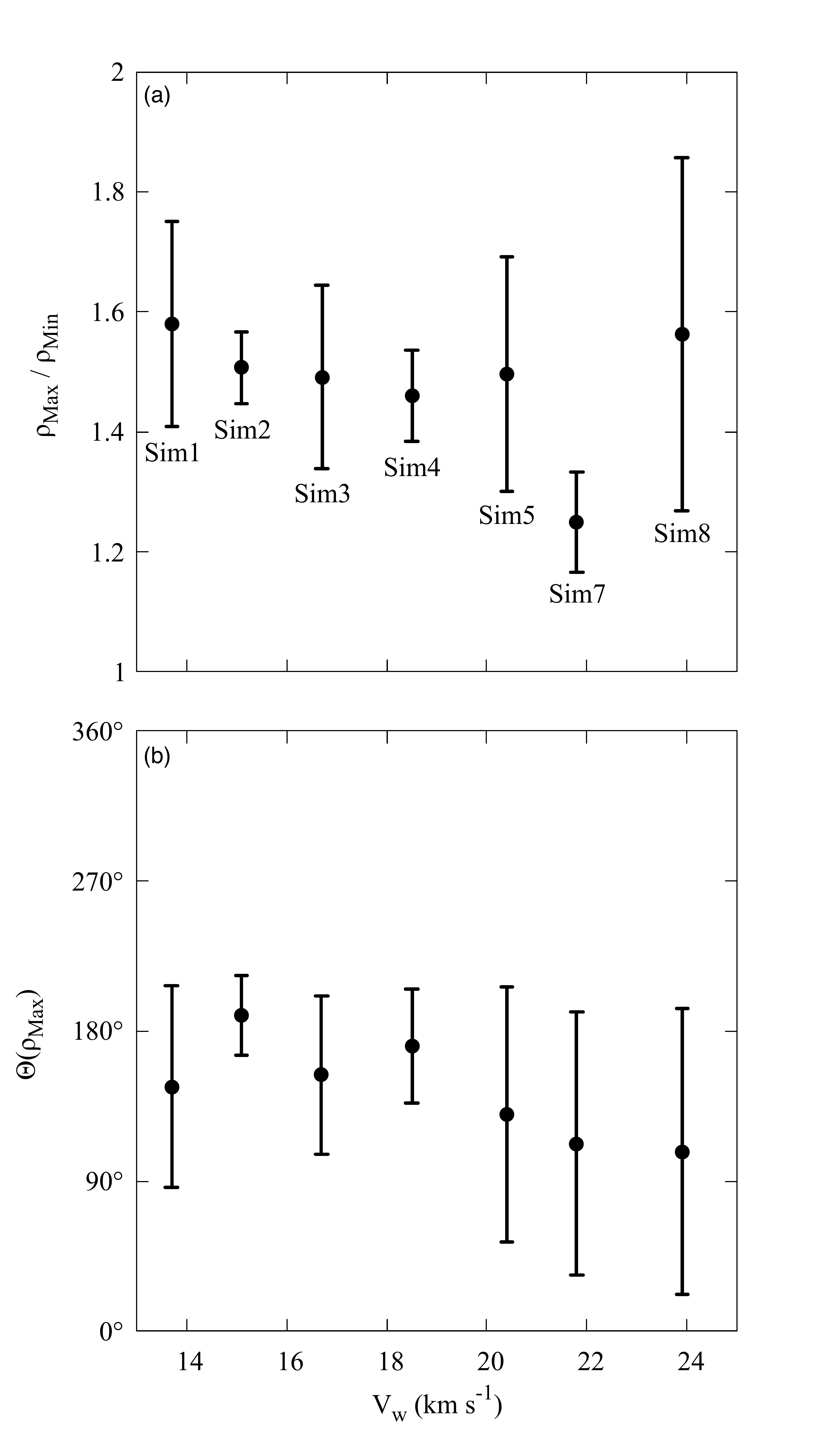}
  \caption{\label{fig:allsimulation}
    (a) Ratio of the density maximum and minimum across azimuthal angle and (b) the angular position $\Theta(\rho_{\rm max})$ of the density peak in the disk as a function of $V_{\rm w}$. In both (a) and (b), the data points shown in filled circles represent the time-averaged value at the last orbit.}
\end{figure}

We investigate the model dependence of the degree of asymmetry, defined as the ratio between the maximum and minimum of the radially-averaged density profile along the azimuthal angle of the disk, $\rho_{\rm Max}/\rho_{\rm Min}$. In the upper panel of Figure~\ref{fig:allsimulation}, the filled circle presents the mean value of time variation of this quantity during the last orbit of simulation, and the error bar shows its standard deviation. The resulting density ratio is nearly independent of the wind models, being $\sim1.5$ on average.

The large temporal variation in the degree of density asymmetry, as indicated by the error bar, has an implication for an observation at a marginal spatial resolution: a time variation of emissivity from the dominant part of the disk may be a natural consequence and contribute to variability in addition to the pulsation of the giant star, the oscillation during binary orbital period, and the intrinsic variation in the stellar wind \citep[e.g.,][]{richie20, galis15}.

The bottom panel of Figure~\ref{fig:allsimulation} shows that the angular direction for $\rho_{\rm Max}$ tends to be toward the giant star ($\Theta \sim 180\arcdeg$) in the relatively slow wind models (Sim1--Sim4), in which relatively large disks are formed. The deviation from $\Theta \sim 180\arcdeg$ along time is, however, quite large probably due to the turbulent motions of inflowing matter, as expected from the fluctuations of the outer outflowing spiral enveloping the two stars. For the faster wind models (Sim5--Sim8), a more refined numerical approach substantiated with increased computing power should be adopted in order to clarify the physical origin of the large uncertainties. 

\subsection{Kinematics of the Accretion Flow}

In addition to the matter distribution around the WD component, the kinematic property of the accretion flow is essential to properly model the spectroscopic observations. \citet{lee19} investigated the line profile of Raman-scattered \ion{O}{6} features at 6825\,\AA\ and 7082\,\AA\ based on the assumption that the \ion{O}{6} emission region of AG~Dra is characterized by the Keplerian motion around the WD component \citep[e.g.,][]{heo16, heo21}. However, one may naturally expect that the real accretion flow should be much more complicated than a simple Keplerian motion due to the presence of the giant component and contribution from the thermal and turbulent motions.

An analysis of Sim2 reveals that the azimuthal velocity component $V_\phi$ of the accretion stream is about 90\% of the Keplerian velocity $V_{\rm K}$. More specifically at $R=0.135$ au, the azimuthal speed of the accretion flow $V_\phi$ is measured to be 53\ \kmps, where the expected Keplerian speed is $V_{\rm K}=61\ \kmps$ (see Fig.~\ref{fig:a2v9Disk}(f)). The sub-Keplerian rotation is probably attained due to the combined effect of the turbulent component and the pressure gradients associated with the adiabatic equation of state, which act against the gravity of the WD.
The sub-Keplerian accretion flow points out an important caveat in interpreting the spectroscopic observations of symbiotic stars including AG~Dra. That is, one may overestimate the size of the Raman \ion{O}{6} emission region when it is deduced under the simple assumption that the flow is Keplerian. 

For example, in the spectrum of AG~Dra investigated by \citet{lee19}, the separation of the peaks in the double peak profile of Raman-scattered \ion{O}{6} at 6825\,\AA\ was observed to be $\simeq 8.2$\,\AA, which was translated to a Keplerian velocity $V_{\rm K}\simeq 27\,\kmps$ of the main \ion{O}{6} emission region around WD. With the adopted value of $M_{\rm WD}=0.6\,\Msun$, the distance to the \ion{O}{6} emission region from the WD is 0.7~au assuming that the flow is Keplerian. If the accretion flow is sub-Keplerian as the result of the Sim2 model indicates, the \ion{O}{6} emission region would be located at a distance of $r_{\rm OVI}=0.5$~au from the WD, which is smaller by a factor $\sim70\%$ than the value deduced under the Keplerian flow assumption.

However, the disk radius for the Sim2 model is $R_{\rm disk}=0.16$\,au, which is only 1/3 of the above estimate of $r_{\rm OVI}$. This substantial discrepancy may indicate other possibilities for radiative transfer modeling with \ion{O}{6} emission regions and neutral regions more sophisticated than \citet{lee19}. For example, based on the spectra obtained with the \emph{Far Ultraviolet Spectroscopic Explorer}, \citet{young05} proposed that the \ion{O}{6} emission line region is located on the illuminated part of the giant atmosphere. It has also been proposed that the \ion{O}{6} emission region may be found in a slowly expanding region from the hot white dwarf \citep{schmid99}. It should be noted that the radiative transfer of Raman-scattered \ion{O}{6} \rev{by \citet{lee19}} was based on an additional assumption that all the neutral matter is present near the giant star. Furthermore, \citet{schmid96} proposed that the kinematics of the neutral wind from the giant component in symbiotic stars is very important in the line formation of Raman-scattered \ion{O}{6} features, pointing out the receding part of the neutral wind contributing to the red enhanced line profile exhibited in Raman \ion{O}{6} features.

\section{Summary and discussion}\label{sec:discussion}

In this work, we have carried out a hydrodynamical study of the stellar wind accretion process in a representative S-type symbiotic star in order to investigate the physical properties of the accretion disk. The {\tt FLASH} code was used incorporating the radiative cooling and an accretion sink. The binary model considered in this work consists of a WD with $M_{\rm WD}=0.6\ \Msun$ and a mass losing giant of $M_{\rm G}=1.5\ \Msun$ with a mass loss rate of $\dot M_{\rm G}=10^{-7}\ \Mspy$. The binary separation is assumed to be 2~au, for which the orbital period is 1.96~yr. These binary parameters are chosen in order to obtain detailed physical properties of the S-type symbiotic star AG~Dra, in which Raman-scattered O~VI lines exhibit double-peak profiles indicative of an accretion flow around the WD component.

In our fiducial Sim2 model, the accretion disk is in a flared shape characterized by the physical dimensions of $R_{\rm disk}=0.16$~au and $H_{\rm disk}=0.03$~au, and the accretion flow is sub-Keplerian with the speed about 90\% of the Keplerian speed. The disk mass reaches $M_{\rm disk} \sim 5\times10^{-8}\ \Msun$ with the temperature of $\sim 6300$~K. 
The accretion rate is $\dot M_{\rm acc} \sim 1.1\times10^{-8}\ \Mspy$. This accretion rate is comparable to the rate of $3.2\times10^{-8}\ \Mspy$ suggested by \citet{greiner97} as a source for hydrogen burning on the surface of WD satisfying the observed X-ray emission flux of AG~Dra in its quiescent state.

In our fast wind models (Sim5--Sim8) with the wind velocity $V_w$ faster than the orbital velocity $V_{\rm WD}$ of the WD, the accretion efficiency $\beta$, the disk radius $R_{\rm disk}$, and the shape of accretion column attached to the WD as a part of the outflowing spiral are in accordance with the BHL theory. The relatively slower Sim1--Sim3 models, however, have the accretion efficiencies considerably higher than the BHL prediction, which is likely related to the presence of detached bow shock preserving the yet-rotating material above the disk radius from ram pressure stripping due to the stellar wind.

The physical properties of the accretion disk formed through capture of the slow stellar wind from the giant donor are quite different from those of the geometrical thin and optically thick accretion disk found in cataclysmic variables resulting from Roche lobe overflow, in which, for example, the accretion flow is almost Keplerian. \citet{perets13} pointed out the similarity of surface density and temperature profiles between wind-fed disks and low-mass protoplanetary disks. A detailed discussion on the hydrodynamical nature of wind-fed disks can be found in the works of e.g., \citet{deval17} and \citet{huarte13}, whereas in this work, we focus on the kinematics and density distribution found in the accretion flow. The temperature distribution in the disk would be affected by the nature of coolants as well as the radiation from the hot WD, which will be the scope of our future work.

The matter distribution in the accretion flow is azimuthally asymmetric, possibly related to the behaviours of streamlines consisting the two inflowing spiral features within the disk. The asymmetric density distribution of the disk matter sustains over a few orbital periods until the end of the numerical simulations. It is found that one side is denser than the opposite side by a factor $\sim1.5$, implying that the local emissivity can be twice stronger than that on the opposite side.

In our future work, we will apply the photoionization computation to the current hydrodynamical results in order to obtain the local emissivities of \ion{O}{6} $\lambda$1032 and $\lambda$1038 and the distribution of neutral hydrogen. The emissivities of \ion{O}{6} will be used as input information for the investigation of the line formation of Raman-scattered \ion{O}{6} features through the Monte Carlo technique adopted by \citet{csj20}.
It remains to be seen that an asymmetric density distribution in the accretion flow is responsible for the multiple-peak profiles often displayed by Raman-scattered \ion{O}{6} features at around 6825\,\AA\ and 7082\,\AA. 

The density affects the flux ratio of \ion{O}{6} $\lambda\lambda$ 1032 and 1038 in such a way that the flux ratio $F_{1032}/F_{1038}$ decreases from 2 to 1 as the density increases. Therefore, the local variation of \ion{O}{6} density leads to varying flux ratio $F_{1032}/F_{1038}$. In particular, the flux ratio $F_{1032}/F_{1038}$ is lower on the side with high density than on the opposite side, which may give rise to the disparity of the line profiles of Raman \ion{O}{6} features at 6825\,\AA\ and 7082\,\AA. One important difficulty to obtain the flux ratio $F_{1032}/F_{1038}$ is possible interstellar extinction of \ion{O}{6} $\lambda$1038 by molecular hydrogen \citep[e.g.][]{schmid99, birriel00}. It is hoped that useful insights into the complex nature of Raman \ion{O}{6} features
are provided through combined studies of hydrodynamics and photoionization modeling.

\acknowledgements
We thank the referee for the constructive comments to help improving the manuscript. 
We are grateful to Seok-Jun Chang, Jeong-Gyu Kim and Kwang-Il Seon for their useful comments.
This research was supported by the National Research Foundation of Korea(NRF) grant funded by the Korea government(MSIT) (No.~NRF-2018R1D1A1B07043944 and NRF-2021R1A2C1008928).
This work was partly supported by the Korea Astronomy and Space Science Institute grant funded by MSIT (Project No. 2022-1-840-05).
The numerical simulations presented here have been performed using the computing resources of a high-performance computing
cluster at the Korea Astronomy and Space Science Institute (KASI).

\begin{table*} 
  \movetableright=1cm
  \caption{\label{tab:repmod}
    Fixed model parameters adopted in the hydrodynamic simulations
    in this work.}
  \begin{tabular}{l|l|l}
    \tableline
    Parameter & Value & Description \\
    \tableline
    $M_{\rm G}$ & 1.5\,\Msun & Mass of the giant star\\
    $\dot M_{\rm G}$ & $10^{-7}$\ \Mspy & Mass-loss rate of the giant star\\
    $M_{\rm WD}$ & 0.6\,\Msun & Mass of the white dwarf\\
    $a$ & 2 au & Binary separation\\
    $r_*$ \& $r_{\rm soft, G}$ & 0.25 au
    & Photospheric and gravitational softening radii of the giant star\\ 
    $T_{\rm eff}$ & 4,000 K & Effective temperature of the giant star\\
    $f$ & 1 & Acceleration factor\\
    \tableline
  \end{tabular}
\end{table*}

\begin{table*} 
  \movetableright=1cm
  \caption{\label{tab:totmodel}
    Model parameters adopted in the eight simulations in this work. The intrinsic wind velocity $V_{\rm w}$ is measured at the distance of 2~au in the corresponding single-star simulation with an ejection of the stellar wind of velocity $V_{\rm ej}$ at the photospheric radius $r_*$ of the giant star. The radius of the accretion sink sphere is denoted by $r_{\rm sink}$ and the softening radius $r_{\rm soft, WD}$ is set to be the same as $r_{\rm sink}$. By $\Delta x_{finest}$, we denote the length of the finest grid cell in the adaptive mesh refinement scheme. For the seven cases at which the accretion disk is defined (i.e., except for Sim6; see Section~\ref{sec:Formation}), we list the radius $R_{\rm disk}$, height $H_{\rm disk}$, and mass $M_{\rm disk}$ that are obtained by averaging over the period at the last orbit of the binary. The standard deviations of the lengths and the mass are less than 10\% of the values.}
  \begin{tabular}{c||c@{ [}c@{]\quad}cc|ccc}
    \tableline
    Model & $V_{\rm w}$ & $V_{\rm ej}$ & $r_{\rm sink}\ \&\ r_{\rm soft, WD}$
    & $\Delta x_{finest}$ & $R_{\rm disk}$ & $H_{\rm disk}$ & $M_{\rm disk}$\\ 
    & \multicolumn{2}{l}{(\kmps)} & (au)
    & ($10^{-3}$ au) & (au) & (au) & ($10^{-8}\,\Msun$)\\
    \tableline
    Sim1  & 13.7 & 6  & 0.05   & 6.25  & 0.14 & 0.03 & 3.5\\
    Sim2  & 15.1 & 9  & 0.05   & 6.25  & 0.16 & 0.03 & 4.6\\
    Sim3  & 16.7 & 12 & 0.05   & 6.25  & 0.14 & 0.03 & 2.9\\
    Sim4  & 18.5 & 15 & 0.05   & 6.25  & 0.15 & 0.03 & 1.5\\
    Sim5  & 20.4 & 18 & 0.05   & 6.25  & 0.12 & 0.02 & 0.4\\
    Sim6  & 21.8 & 20 & 0.05   & 6.25  &  -   &  -   &  - \\
    Sim7  & 21.8 & 20 & 0.025  & 6.25  & 0.11 & 0.02 & 1.0\\
    Sim8  & 23.9 & 23 & 0.0125 & 3.125 & 0.06 & 0.01 & 0.1\\
    \tableline
  \end{tabular}
\end{table*}

\end{document}